\documentclass[12pt]{article}

\usepackage{epsfig,multicol,multirow,cite}
\usepackage{amsmath,subfigure,latexsym,amssymb}
\usepackage[figuresright]{rotating}
\usepackage{tikz, graphicx}

\usepackage[colorlinks=true,citecolor=magenta,urlcolor=blue]{hyperref}
\usepackage{hyperref}

\newcommand{\be}{\begin{equation}}
\newcommand{\ee}{\end{equation}}
\newcommand{\nn}{\nonumber}
\newcommand{\bea}{\begin{eqnarray}}
\newcommand{\eea}{\end{eqnarray}} 

\newcommand{\la}{\langle}
\newcommand{\ra}{\rangle}

\newcommand{\Z}{\mathbb{Z}}
\newcommand{\R}{{\kern+.25em\sf{R}\kern-.78em\sf{I} \kern+.78em\kern-.25em}}
\newcommand{\RR}{{\kern+.25em\sf{R}\kern-.6em\sf{I} \kern+.6em\kern-.25em}}
\newcommand{\N}{{\kern+.25em\sf{N}\kern-.78em\sf{I} \kern+.78em\kern-.25em}}
\newcommand{\C}{{\kern+.25em\sf{C}\kern-.50em\sf{I} \kern+.50em\kern-.25em}}

\newcommand{\corresp}{\stackrel{\wedge}{=}}

\newcommand{\iso}{\ \raisebox{0.5ex}{\ensuremath{\stackrel{\sim}{\rule{12pt}{0.7pt}}}\ \,}}

\usepackage{color}
\definecolor{col_red}{rgb}{1.0,0.0,0.0}

\makeatletter
\@addtoreset{equation}{section}
\makeatother

\begin{document}
 
\begin{center}
{\Large\bf Phase diagram of chiral 2-flavor QCD}

\vspace*{6mm}

{\Large\bf based on an effective approach} \\

\vspace*{1cm}

Edgar L\'{o}pez-Contreras,
Jos\'{e} Antonio Garc\'{\i}a-Hern\'{a}ndez,\vspace*{1mm} \\
El\'{\i}as Natanael Polanco-Eu\'{a}n,
Wolfgang Bietenholz
\\
\ \\
Instituto de Ciencias Nucleares \vspace*{1mm} \\
Universidad Nacional Aut\'{o}noma de M\'{e}xico \vspace*{1mm} \\
A.P.\ 70-543, C.P.\ 04510 Ciudad de M\'{e}xico, Mexico\\

\end{center}

\vspace*{6mm}

\noindent
Despite intense experimental and theoretical research, the
QCD phase diagram at finite baryon density remains to a large
extent unexplored.
From the theoretical side, the obvious non-perturbative approach
is lattice QCD simulations, which are, however, obstructed
by a severe sign problem. Here we employ the O(4) non-linear
$\sigma$-model as an effective theory for 2-flavor QCD in the chiral
limit. The identical pattern of spontaneous symmetry breaking
indicates that they belong to the same universality class.
We assume high temperature dimensional reduction to the
3d O(4) model, with topological charge taking the role
of the baryon number, along the lines of Skyrme's model.
In this effective formulation,
the baryonic chemical potential $\mu_{B}$ can be
included in the lattice formulation without causing
any sign problem in Monte Carlo simulations.
This allows us to pin down the critical line, {\it i.e.}\
the critical temperature $T_{\rm c}(\mu_{B})$, which decreases
monotonically for increasing $\mu_{B}$.
In the range $0 < \mu_{B} \lesssim 309~{\rm MeV}$
and $132~{\rm MeV} \gtrsim T_{\rm c} \gtrsim 106~{\rm MeV}$, we do not
find a Critical Endpoint (CEP), although there are hints for it to be
in the vicinity of the maximal $\mu_{B}$-value that we could explore.

\newpage

\tableofcontents

\section{The mysterious QCD phase diagram}

\label{intro}
While a large part of the particle physics community discusses
(or speculates) about physics beyond the Standard Model,
there are still interesting open questions within the
Standard Model. A particularly outstanding issue is the
{\em QCD phase diagram.}

At considerable baryon density, which is especially
relevant for neutron stars and for the early
Universe, this phase diagram is still a matter of
speculations. It is widely assumed --- though not certain ---
that the cross-over between the confined and deconfined
phase at low baryon density, which corresponds
to a small baryonic chemical potential $\mu_{\rm B}$, ends
in a Critical Endpoint (CEP). A number of laboratories are
performing experimental searches for this CEP, or are planning
to do so: this includes the Relativistic Heavy Ion Collider (RHIC),
SPS and all major LHC experiments at CERN, the
Nuclotron-based Ion Collider fAcility (NICA) in Dubna,
the Facility for Antiproton and Ion Research (FAIR) in Darmstadt,
the Japan Proton Accelerator Research Complex (J-PARC) and
the Heavy Ion Research Facility (HIRFL) in Lanzhou.

At $\mu_{\rm B} = 0$, the pseudocritical temperature $T_{\rm pc}$ has
been successfully measured in lattice QCD simulations. This is
the obvious method to be applied, since the issue is clearly
non-perturbative. Due to the
nature of the cross-over, the value of $T_{\rm pc}$ that one obtains
depends somewhat on the criterion that one refers to. There is
a consensus that $T_{\rm pc} \approx 155 \, {\rm MeV}$ is a sensible
value \cite{BazavovBhattacharya}.

In the theoretical framework,
we can also study the scenario of vanishing masses of the light quark
flavors, $m_{u}=m_{d}=0$. In many respects, this is a good approximation
to reality, {\it e.g.}\ the nucleon mass is just reduced at
percent-level \cite{chiralLQCD}.
In the chiral limit, the cross-over turns into a second order phase
transition \cite{chiralPT}, and the critical temperature at
$\mu_{\rm B} = 0$ was obtained at $T_{\rm c} \simeq 132 \, {\rm MeV}$
\cite{Ding19} with the $s$-quark at its physical mass, and
similarly at $T_{\rm c} \simeq 134 \, {\rm MeV}$ when even the
$c$-quark is included \cite{KLT21}. We take the former value as
our reference for the $T_{\rm c}$. An alternative study in Ref.\
\cite{KoSin06} actually simulated chiral 2-flavor QCD, with
staggered fermions and an irrelevant 4-fermion term, though
with less stringent results for the critical temperature.

In this work, we are going to refer to chiral 2-flavor QCD. In
this theory, the second-order phase transition line
(which is free of ambiguities) is again expected
to end in a CEP, where the phase transition turns into
first order. In the region of even larger $\mu_{\rm B}$, one may
speculate about a variety of new phases, as in the case of
physical quark masses.

The reason why the extension of the QCD phase diagram
to finite $\mu_{\rm B}$ has not yet been explored by lattice
simulations is that they are confronted with a severe sign
problem. The Euclidean action becomes complex, hence it
cannot be used to directly define a probability density for the
importance sampling of the gauge configurations. In principle, the
imaginary part can be treated by re-weighting.
However, for that approach to provide results with reasonably
small errors, tremendous statistics are necessary --- the
statistical requirement grows exponentially with the volume.

Attempts to overcome this sign problem are reviewed for instance
in Refs.\ \cite{Philippe}. They include simulations with a complex
Langevin algorithm (where, however, the compact link variables
leave the gauge group SU(3)), simulations at imaginary $\mu_{\rm B}$
(one then tries to extrapolate $\mu_{\rm B}^{2}$ from negative
to positive values), or the numerical computation of Taylor series
coefficients at $\mu_{\rm B}=0$ as another basis for an extrapolation
to $\mu_{\rm B} >0$.

So far no breakthrough has been achieved with such approaches,
hence for the time being it is fully legitimate to employ effective
models for this study. A number of such models have been
investigated already, such as the Nambu-Jona-Lasinio
\cite{NJL} and Polyakov-Nambu-Jona-Lasinio model
\cite{PNJL}, the linear $\sigma$-model \cite{LSM},
as well as a holographic image approach \cite{holography}.

In this work, our effective theory is the O(4) non-linear
$\sigma$-model, in the spirit of Skyrme's model \cite{Skyrme}.
The concept simplifies if we assume dimensional reduction to
the 3d O(4), as we are going to discuss in Section \ref{model}.
We anticipate that the emerging topological charge
$Q$ corresponds to the baryon number $B$, and that this
formulation enables lattice simulations at baryonic chemical
potential $\mu_{B} >0$ without any sign problem. Preliminary
results of this work have been presented in a thesis \cite{Edgar}
and in two proceedings contributions \cite{procs}.

Section \ref{model} motivates our effective theory, and Section
\ref{lattice} explains the lattice formulation that we use,
in particular regarding the topological charge. Section \ref{critical}
presents our numerical results, which are further discussed ---
and compared to other conjectures in the literature --- in
Section \ref{conclu}.

\section{The O(4) model as an effective theory for chiral 2-flavor QCD}

\label{model}
In Nature, two quark flavors are two orders of magnitude lighter
than the intrinsic scale of QCD, $m_{u}, \, m_{d} \ll \Lambda_{\rm QCD}$,
hence the chiral limit $m_{u} = m_{d} = 0$ is often a good approximation,
as we mentioned before. In this limit, the left- and right-handed
quarks  of 2-flavor QCD decouple, and the Lagrangian has a global
${\rm U}(2)_{\rm L} \otimes {\rm U}(2)_{\rm R}$ symmetry, where the
subscripts denote the quark chiralities. If we separate the two
U(1) phase factors (one of them accounts for baryon number conservation
and the other one breaks explicitly under quantization due to the axial
anomaly), we observe --- at low energy --- the spontaneous breaking
of chiral flavor symmetry as follows, 
$${\rm SU}(2)_{\rm L} \otimes {\rm SU}(2)_{\rm R} \ \to \
{\rm SU}(2)_{\rm L=R} \ . $$
Adding quark masses turns the three emerging Nambu-Goldstone bosons
into the massive pion triplet, but here we stay in the massless case.

The low-energy effective theory is formulated
in terms of the Nambu-Goldstone bosons, with a Lagrangian
that is invariant under the unbroken symmetry group.
We represent the Nambu-Goldstone field by the classical spin
$\vec e(x) \in S^{3}$ ({\it i.e.}\ $\vec e (x) \in \R^{4}$ and
$|\vec e (x)| =1, \ \forall x$). We restrict the effective
Lagrangian in continuous Euclidean space to the leading term
\be
   {\cal L}_{\rm eff} = \frac{F_{\pi}^{2}}{2} \, \partial_{\mu}
   \vec e (x) \cdot \partial_{\mu} \vec e (x) \ ,
\ee
which has a global O(4) symmetry. $F_{\pi}$ is a low-energy constant,
namely the pion decay constant, which has the phenomenological value
$F_{\pi} \simeq 92.4 \ {\rm MeV}$. We do not include higher-derivative
terms, nor an explicit symmetry-breaking term $-\vec h \cdot \vec e(x)$,
where the norm of the (constant) external ``magnetic field'',
$|\vec h|$, would give rise to a degenerate quark mass, $m_{u}=m_{d}$.

In its absence (and in infinite volume),
the symmetry breaking ${\rm O}(4) \to {\rm O}(3)$
is spontaneous; it can be viewed as a spontaneous ``magnetization''.
Thus we have local isomorphies (which we denote by the symbol
$\iso$) both before and after symmetry breaking,
\be
\{ \ {\rm SU}(2)_{\rm L} \otimes {\rm SU}(2)_{\rm R}
\iso
{\rm O}(4)
\ \} \to \{ \ {\rm SU}(2)_{\rm L=R} \iso
{\rm O}(3) \ \} \ .
\ee
If the O(4) model and chiral 2-flavor QCD are both formulated in
4 dimensions, this relation strongly suggests that their critical
points belong to the same universality class.
The field $\vec e (x) \in S^{3} \mathrel{\widehat{=}}
{\rm O}(4)/{\rm O} (3)$
is formulated in the coset space of the symmetry breaking, hence it
represents the (massless) pions in the broken phase, according to
Chiral Perturbation Theory. The question remains how this mesonic
model can address baryons.

Inspired by Skyrme's model \cite{Skyrme}, but following a somewhat
different approach, we assume the temperature $T$ to be
high enough for the system to undergo dimensional reduction.
This means that the configurations $[\vec e \, ]$, which dominate
the functional integral, are approximately constant in the Euclidean
time direction: its extent $\beta$ is short, so the periodic boundary
conditions suppress the temporal non-zero modes, and the action (in
a spatial volume $V = L^{3}, \ L \gg \beta$) can be approximated as
\bea
S[\vec e \, ] &=& \int_{0}^{\beta} dt_{\rm E} \int_{V} d^{3}x \
\frac{F_{\pi}^{2}}{2} \, \partial_{\mu}
\vec e (x) \cdot \partial_{\mu} \vec e (x) \nn \\
&\simeq & \beta \int_{V} d^{3}x \ \frac{F_{\pi}^{2}}{2} \,
\partial_{i} \vec e (x) \cdot \partial_{i} \vec e (x)
= \beta H[\vec e \, ] \ ,
\eea
where $i=1,2,3$, and $H$ is the Hamilton function.\footnote{Here,
$\beta$ represents the inverse temperature, $\beta = 1/T$, {\it i.e.}\
we refer to units where the Boltzmann constant is 1. In lattice units,
we are going to deal with values $\beta_{\rm lat} \approx 1 \ll L$,
see Sections \ref{lattice} and \ref{critical}.}

The important property is that the 3d O(4) model
--- with periodic boundary conditions ---
has topological sectors, due to $\pi_{3}(S^{3}) = \Z$, as all
$(N-1)$-dimensional O($N$) models. Hence we can assign to each
continuous configuration a winding number, or topological charge,
$Q[\vec e\, ] \in \Z$, so the set of continuous configurations
is divided into topological sectors. Thanks to the assumption
of dimensional reduction, we do not need Skyrme's ``stabilizing
term'' \cite{Skyrme}.

Still, we follow Skyrme and other authors \cite{Skyrme,Skyrme2}
by identifying the topological charge $Q$ with the baryon number
$B$. This can be justified by considering the corresponding currents,
and that is the way how the mesonic model manages to take baryons
into account. In this formulation, the baryonic chemical potential
$\mu_{B}$ takes the role of an imaginary vacuum angle $\theta$,
which leaves the Hamilton function real,
\be  \label{hami}
 H[\vec e \, ] = \int_{V} d^{3}x \ \frac{F_{\pi}^{2}}{2} \,
\partial_{i} \vec e (x) \cdot \partial_{i} \vec e (x) -
\mu_{B} Q[\vec e \,] \in \R \ .
\ee
Thus this model represents chiral 2-flavor QCD at finite
baryon density, by a well-motivated effective theory, which can
be simulated on the lattice without any sign problem, as we
are going to discuss next.

\section{Lattice formulation and simulation}

\label{lattice}
\subsection{Lattice action}

In order to simulate the 3d O(4) model, we need to specify a
lattice formulation. We are going to use cubic lattices
with volumes $V = L^{3}$, always with periodic boundary
conditions in all directions. We first use lattice units
(lattice spacing 1), before finally employing the critical
temperature $T_{\rm c}$ to convert the parameters and results to
physical units.

For the Euclidean lattice action $S_{\rm lat}$, we use the
standard regularization,
\bea
\frac{1}{2} \, \partial_{i} \vec e(x) \cdot \partial_{i} \vec e(x)
& \to & \frac{1}{2} \, (\vec e_{x + \hat i} - \vec e_{x})^{2}
= - \vec e_{x} \cdot \vec e_{x + \hat i} + {\rm const.} \nn \\
\Rightarrow \ S_{\rm lat}[\vec e \, ] &=& -\beta_{\rm lat}
\Big( \sum_{\langle xy \rangle} \vec e_{x} \cdot \vec e_{y}
+ \mu_{B,{\rm lat}} Q [\vec e \, ] \Big)
= \ \beta_{\rm lat} H_{\rm lat}[\vec e \, ] \ , \qquad 
\label{Hlat}
\eea
where $i=1,2,3$ is summed over, $\hat i$ is the lattice unit vector
in $i$-direction, and $\langle xy \rangle$ denotes nearest-neighbor
lattice sites.

\subsection{Topological charge}

The challenge is the definition and computation of the topological
charge $Q[\vec e \, ]$. All lattice configurations can be continuously
deformed into one another, so topological sectors do not exist
{\it a priori}. Any formulation of $Q[\vec e \, ]$ implies some
kind of continuum interpolation, which can be ambiguous
(in particular for rough configurations).

As a guide-line, we follow the geometric definition advocated
by Berg and L\"{u}scher in the 2d O(3) model \cite{BergLuscher}.
The idea is to split the lattice plaquettes into triangles and
compute, for each triangle, the solid angle of the spherical
triangle (with minimal area) that the three spins at its
vertices span on $S^{2}$. If we fix a consistent orientation of
all triangles (such that the edges between adjacent triangles
contribute in the opposite direction), and attach a sign to the
spherical triangle $A_{xyz}[\vec e\,]$ depending on its orientation,
their sum --- normalized by the area of $S^{2}$ --- must be an
integer. This is a sensible definition of the topological charge $Q$,
\be
Q[\vec e\,] = \frac{1}{4 \pi} \sum_{\langle xyz\rangle}
A_{xyz} [\vec e \, ] \in \Z \ .
\ee
The sum runs over all index triplets which are vertices of the
same triangle. In particular, the property that this definition
assigns an integer value to each lattice configuration is a
great advantage of the geometric definition compared to other
suggestions in the literature.
(We can ignore the exceptional cases where the spherical triangle
with minimal area $|A_{xyz}|$ is ambiguous, since these
configurations form a subset of measure zero.) 

\begin{figure}[h!]
\begin{center}
\hspace*{-1cm}
\includegraphics[angle=0,width=.3\linewidth]{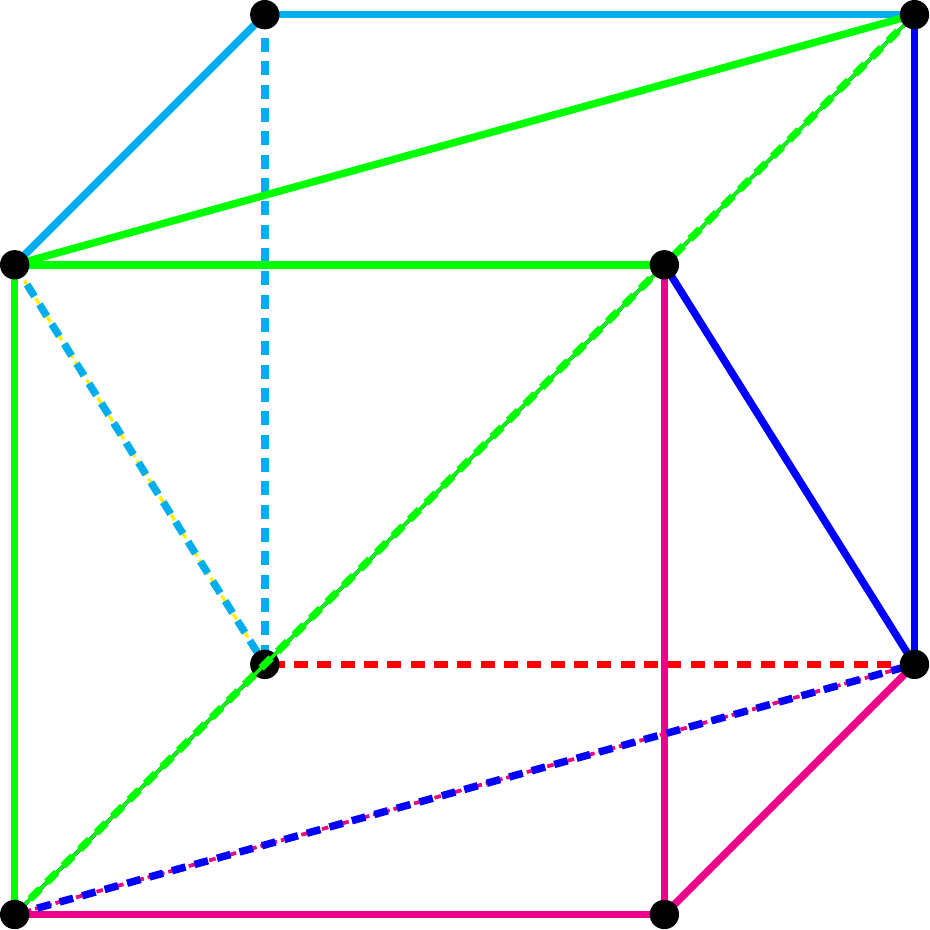}
\hspace*{1cm}
\includegraphics[angle=0,width=.32\linewidth]{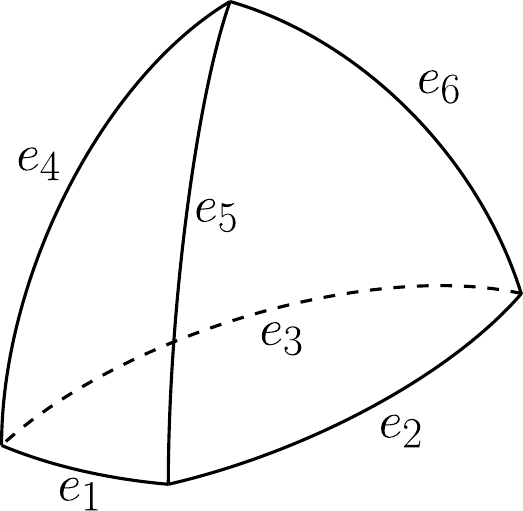}
\end{center}
\caption{Left: Illustration of the division of a lattice unit
cube into six tetrahedra (with distinct colors and line-types),
as used in this work.
Right: A schematic sketch of a spherical tetrahedron on the sphere
$S^{3}$, with edges $e_{1}, \dots ,e_{6}$.}
\label{tetrafig}
\end{figure}

In the 3d O(4) model, we follow the same concept, but it becomes
more complicated. We decompose each lattice unit cube into six
tetrahedra, with an orientation such that the faces between adjacent
tetrahedra contribute with opposite planar orientations \cite{UJW}.
A way to do so is depicted in Figure \ref{tetrafig} on the left.

The volumes and shapes of these tetrahedra differ, but what matters
is that they are space-filling. (A decomposition of the unit
cubes into five tetrahedra would also be possible, as
depicted in Ref.\ \cite{Senechal}.)\footnote{In the 2d O(3)
model, it is an elegant alternative to use a lattice
consisting of regular triangles, which the oriented solid angles
$A_{xyz}$ refer to. Hence one might be tempted to suggest for the
3d O(4) model a lattice of regular tetrahedra. This fails,
however, since these tetrahedra are not exactly space-filling.
The fascinating history of the analysis of space-filling tetrahedra,
which began in Ancient Greece, is reviewed in Ref.\ \cite{Senechal}.}

The four spins at the vertices $\la wxyz\ra$
of any tetrahedron span a spherical
tetrahedron on $S^{3}$, as symbolically illustrated in the right
panel of Figure \ref{tetrafig}. We again refer to the one with minimal
volume, $|V_{wxyz}|$, and attach a sign factor depending on the
orientation of this tetrahedron (it corresponds to the sign of
the determinant of the matrix
$(\vec e_{w},\vec e_{x},\vec e_{y},\vec e_{z})$).
Summing up these oriented volumes, and normalizing by the volume of
$S^{3}$, leads to the geometrically defined topological lattice charge
\be  \label{Qtop}
Q[\vec e \, ] = \frac{1}{2 \pi^{2}} \sum_{\la wxyz\ra} V_{wxyz}[\vec e \, ]
\in \Z \ ,
\ee
where the sum runs over all lattice site quadruplets that
represent vertices of the same tetrahedron.

This conceptual extension of the geometric charge to the
3d O(4) model seems straightforward, but the computation of
$V_{wxyz}[\vec e \, ]$, {\it i.e.}\ of the oriented volume of
a spherical tetrahedron, is not. It was only in 2012 that
Murakami published a paper with two mathematical recipes
(procedures rather than just formulae) for this purpose
\cite{Murakami}.

The numerical implementation has been successfully achieved
by Nava Blanco \cite{Nava} and also by two of the present
authors \cite{Edgar,JAGH}. The property (\ref{Qtop})
(with an integer to machine precision) for arbitrary configurations
is a highly sensitive testing criterion (it fails, for instance,
if the orientations are not handled in a fully consistent way).

Moreover, since $Q[\vec e \,]$ represents a winding number, it can also
be computed by fixing an arbitrary reference point on $S^{3}$ and counting
--- in an oriented way --- how many spherical tetrahedra contain
this point. This consistency test was again successful for numerous
random configurations, and that method is numerically faster.
It does, however, not keep track of the local topological charge
density, which is given by $V_{wxyz}[\vec e \,]/(2 \pi^{2})$.

\subsection{Simulation}
\label{simu}

Our simulations were performed with the multi-cluster algorithm
\cite{SwenWang,Wolff}, and confirmed by consistency tests with
the single-cluster algorithm \cite{Wolff}. Compared to local-update
algorithms, the collective cluster flips strongly suppress the critical
slowing down near the second order phase transition (see below).
The availability of this powerful algorithm is another
benefit of the effective theory used in this work.

At $\mu_{B}=0$, we follow the standard procedure \cite{Wolff}:
we choose a random Wolff direction $\vec r$ in 4d spin space,
$\vec r \in S^{3}$, 
and define a ``flip'' of some spin $\vec e_{x}$
as its reflection at the Wolff hyperplane perpendicular to $\vec r$,
$\vec e_{x} \to \vec e_{x}\,'
  = \vec e_{x} - 2 (\vec r \cdot \vec e_{x}) \, \vec r$.
The energy change under spin flips fixes the probability to set
a ``bond'' between two nearest-neighbor spins.
A set of spins connected by bonds constitutes
a ``cluster''. At last, the spins in each cluster are collectively
flipped with probability $1/2$. Then one chooses a new
Wolff direction and repeats the steps above.

At $\mu_{B} \neq 0$, the bonds are set in the same manner as at
$\mu_{B} = 0$, but the chemical potential alters the cluster-flip
probability, which now turns into a Metropolis accept/reject step,
based on the term $- \mu_{B,{\rm lat}} Q[\vec e\, ]$ in 
$H_{\rm lat}[\vec e\, ]$, see eq.\ (\ref{Hlat}). This way
to include the topological term is analogous to the procedures
described in Refs.\ \cite{Wang}.

\begin{figure}[h!]
\vspace*{-4cm}
\hspace*{2cm}
\includegraphics[angle=0,width=0.95\linewidth]{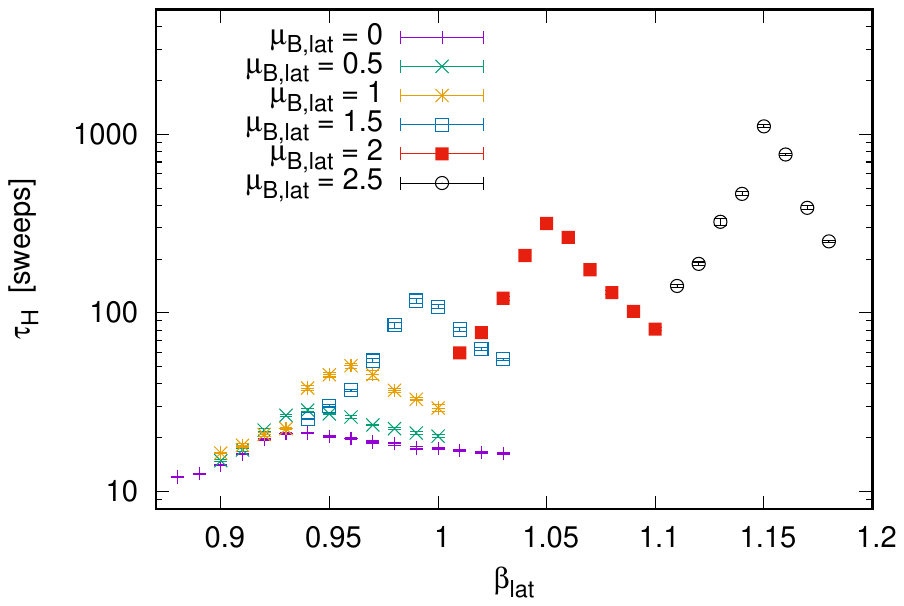} \vspace*{-5.5cm} \\
\hspace*{2cm}
\includegraphics[angle=0,width=0.95\linewidth]{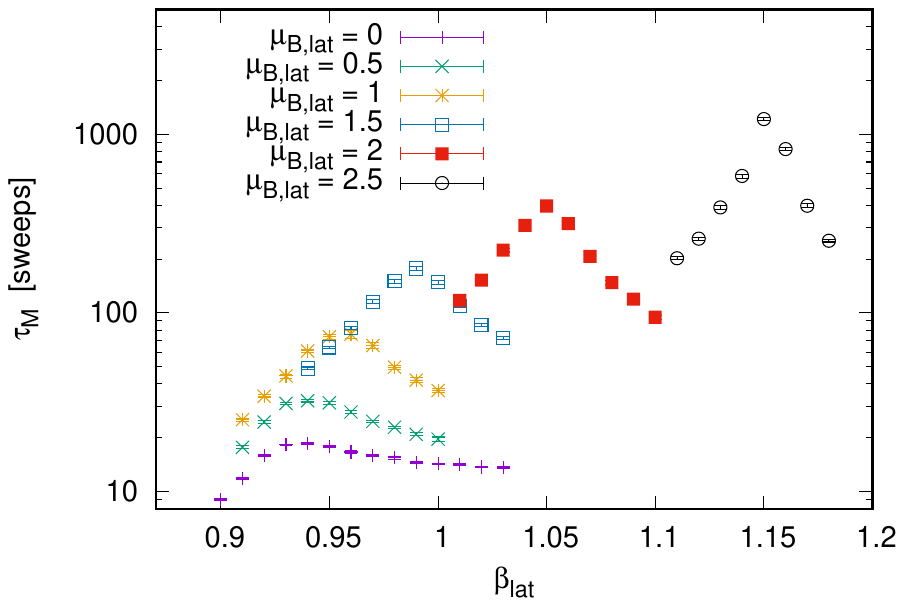} \vspace*{-5.5cm} \\
\hspace*{2cm}
\includegraphics[angle=0,width=0.95\linewidth]{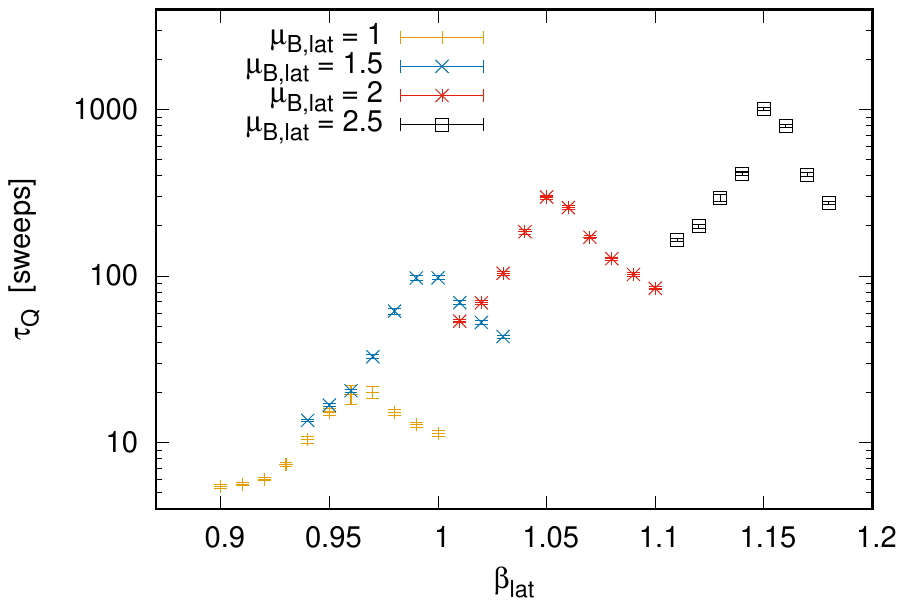} \vspace*{-1.5cm} \\
\vspace*{-8mm}
\caption{The auto-correlation time with respect to the energy,
  $\tau_{H}$ (top), the magnetization, $\tau_{M}$ (center),
  and the topological charge, $\tau_{Q}$ (bottom), all in units of
  multi-cluster update steps (``sweeps''). We show the exponential
  auto-correlation
  time obtained by fits of good qualities, as the small uncertainties
  confirm. (For $\mu_{B,{\rm lat}} <1$, $\tau_Q$ is of ${\cal O}(1)$ sweep.)
  The peak locations provide a first hint for the critical
  value of $\beta_{\rm lat}$, and the peak heights show that the simulations
become rapidly more demanding for increasing $\mu_{B,{\rm lat}}$.}
\label{autocorr}
\vspace*{-7mm}
\end{figure}
Once the chemical potential is turned on, the
auto-correlation time increases rapidly, even for the
cluster algorithm. The reason is that the clusters
are more and more pushed to one side of the Wolff hyperplane. 
Figure \ref{autocorr} shows the auto-correlation times
with respect to the Hamilton function (or energy) $\tau_{H}$,
the magnetization $\tau_{M}$, and the topological charge,
$\tau_{Q}$, in units of multi-cluster updates (which we denote
as ``sweeps'').
We refer to the exponential auto-correlation time $\tau$
obtained by fits of good qualities.
We show data for a variety of baryonic chemical potentials
$\mu_{B,{\rm lat}}$, as functions of the inverse temperature
$\beta_{\rm lat}$, both in lattice units, in lattice volumes
$V = 20^{3}$. \\

From Figure \ref{autocorr}, we infer the following observations:

\begin{itemize}

\item All three auto-correlation times are similar, $\tau_{H} \approx
  \tau_{M} \approx \tau_{Q}$, which is a great virtue compared to local
  update algorithms, like Metropolis, Glauber or heatbath, where it
  is a notorious problem to change the topological sector (except for
  high temperature), such that $\tau_{Q}$ tends to be huge
  close to criticality (``topological freezing'').

\item For $\mu_{B,{\rm lat}} \gtrsim 1$, marked peaks show up. They provide a
  clear hint about the critical temperature. We see that these peaks
  move to lower temperatures as $\mu_{B,{\rm lat}}$ increases, as generally
  expected in QCD.

\item The peak height also increases with $\mu_{B,{\rm lat}}$, and
  at $\mu_{B,{\rm lat}} = 2.5$ it attains auto-correlation times of
  more than 1000 sweeps. We performed all our numerical measurements
  --- to be reported in Section \ref{critical} --- such that they
  were separated by at least $2 \tau_{\rm max}$, where $\tau_{\rm max}$
  is the longest auto-correlation time for the three observables
  under consideration.
  In this manner, the measurements can be considered as
  independent, and the statistical standard error applies.

  This shows that simulations at $\mu_{B,{\rm lat}} = 2.5$ are
  hard. This difficulty prevented us from proceeding
  to larger values of $\mu_{B,{\rm lat}}$ or larger lattice
  sizes than $L=20$. Of course, increasing $L$ does not
  only have a direct computational cost, but also an indirect
  one, because $\tau_{\rm max}$ again increases rapidly.
  Still, our results obtained at $L \leq 20$ are
  conclusive for the critical temperature,
  as we are going to see.
  
\end{itemize}

\begin{figure}[h!]
\begin{center}
\includegraphics[angle=0,width=.45\linewidth]{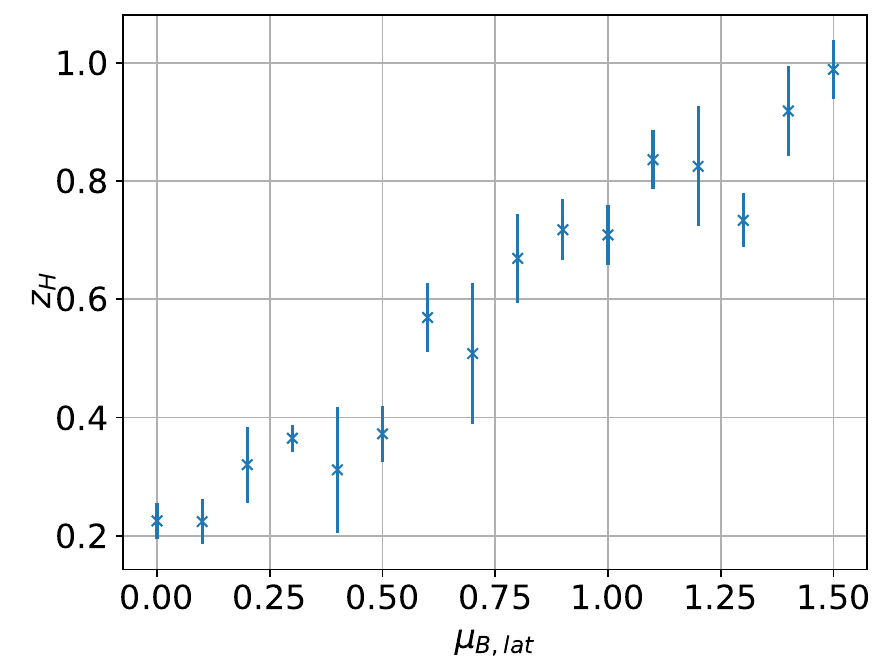}
\includegraphics[angle=0,width=.45\linewidth]{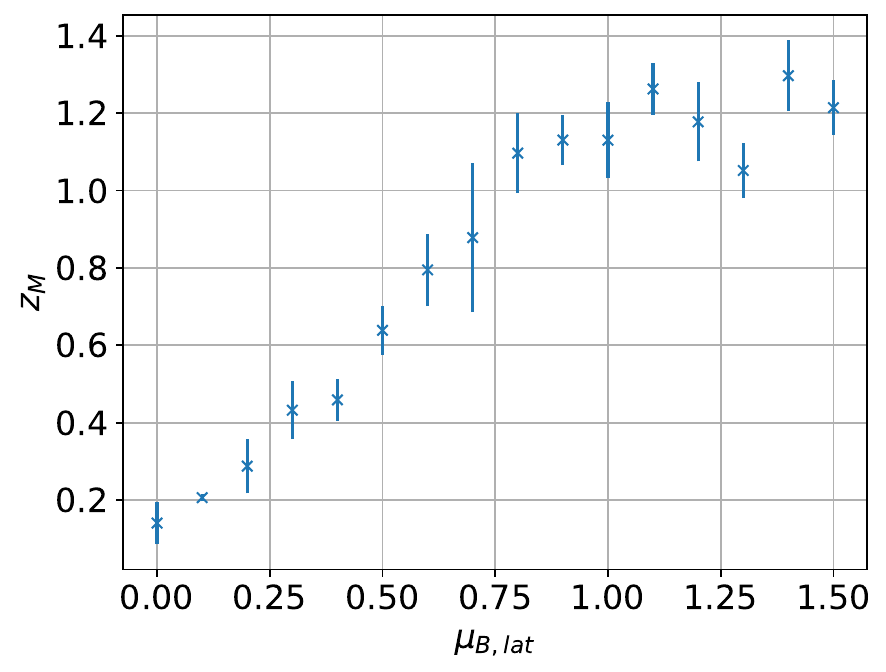}
\includegraphics[angle=0,width=.45\linewidth]{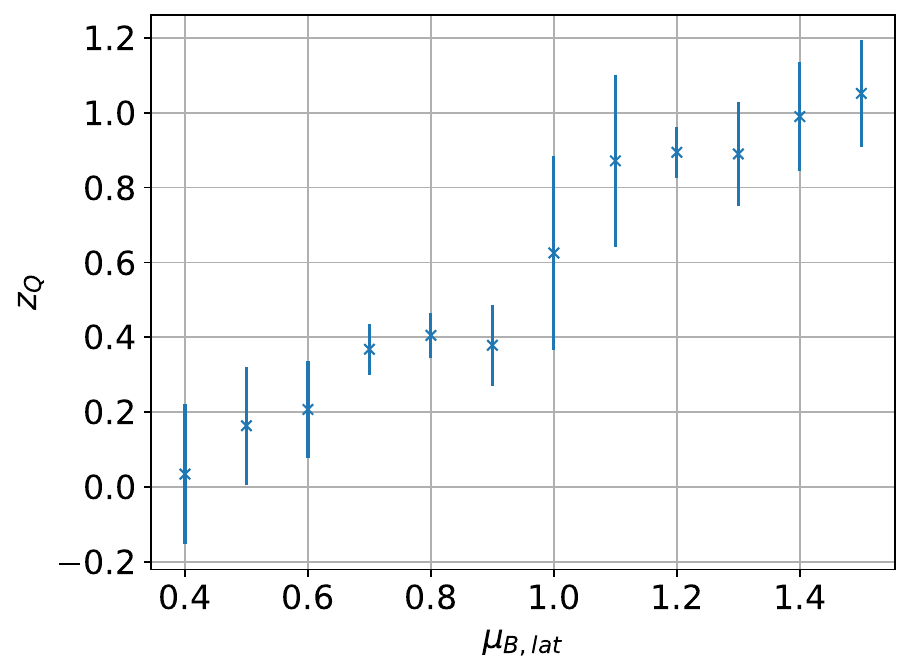}
\end{center}
\vspace*{-6mm}
\caption{The dynamical critical exponent, based on 
  relation (\ref{tauxiz}), with respect to the energy, $z_{H}$,
  the magnetization, $z_{M}$, and the topological charge, $z_{Q}$.
(They are obtained for $\beta_{\rm lat}$ in the vicinity of its critical value,
  to be discussed in Section \ref{critical}.)
  The $z$-values are very low at $\mu_{B,{\rm lat}} \approx 0$, which
  confirms that the critical slowing down is mild in this case, thanks to the
  cluster algorithm. However, when the baryonic chemical potential increases
  to $\mu_{B,{\rm lat}} \approx 1$, the dynamical critical exponents also
  attain values around 1.}
\label{zdyncrit}
\end{figure}

For completeness, let us also consider 
the notorious problem of ``critical slowing down''
in this case. It is quantified by the relation
\be  \label{tauxiz}
\tau_{\cal O} \propto \xi^{z_{\cal O}} \ ,
\ee
where $\tau_{\cal O}$ is the autocorrelation time with respect to some
observable ${\cal O}$, and $\xi$ is the correlation length (we are
going to show results for $\xi$ in Section \ref{critical}).
The proportionality relation refers
to the asymptotic behavior as one approaches criticality, and
$z_{\cal O} > 0$ signals the property of critical slowing down, which
is worse the larger the maximal value of $z_{\cal O}$ for the
observables of interest. For local-update algorithms,
one usually obtains $z_{\cal O} \approx 2$, which poses a serious
problem. Our results (obtained with the cluster algorithm) are shown
in Figure \ref{zdyncrit}: at $\mu_{B,{\rm lat}} \approx 0$, the slowing
down is relatively harmless, but it increases significantly when we
proceed to $\mu_{B,{\rm lat}} \approx 1$.

\section{Critical line and search for a Critical Endpoint}

\label{critical}
We repeat that, in the framework of this effective model, a
degenerate quark mass corresponds to the strength of an external
magnetic field $\vec h$, which gives rise to an additional term
$-\vec h \cdot \sum_{x} \vec e_{x}$ in $H_{\rm lat}[\vec e \, ]$.
Here we deal with $\vec h = \vec 0$, hence we refer to the chiral
limit, where the quark masses and the pion mass vanish. We will,
however, occasionally refer also to derivatives with respect to
$\vec h$, $\partial_{\vec h}$, which are taken at $\vec h = \vec 0$.

We simulate in cubic lattice volumes, $L^{3}$, with $L= 10,\, 12,\,
16,\, 20$, at
$$\mu_{\rm B,lat} = 0, \, 0.1,\, 0.2, \dots , 1.4,\, 1.5; \, 2,\, 2.5 \ .$$
At low $\mu_{\rm B,lat}$, we could proceed to larger volumes,
but the sharp increase in the auto-correlation time prevents us
from doing so at our largest values of $\mu_{\rm B,lat}$, as Figure
\ref{autocorr} shows.

\subsection{Numerical results}
\label{numires}

We measure observables, which are given by first or second derivatives
of the free energy $F = - (\ln Z) / \beta$.
According to Ehrenfest's classification, a discontinuity with
respect to $\beta$ indicates a first or second order phase
transition, respectively. At each value of $\mu_{\rm B,lat}$, we
have to search for the $\beta$-value where a phase transition
takes place, $\beta_{\rm c,lat} (\mu_{B,{\rm lat}})$, and we measure
a set of observables in its vicinity.

All our results to be shown below are based on $10^{4}$ configurations
for each parameter set.
We repeat that they are perfectly decorrelated, {\it i.e.}\ measurements
are separated at least by $2 \tau_{\rm max}$ sweeps (where $\tau_{\rm max}$
is the maximum among $\tau_{H}$, $\tau_{M}$, $\tau_{Q}$),
even at the peaks in Figure \ref{autocorr}, such that the statistical
standard error applies. We first consider a set of results obtained
at $L=20$.

\begin{figure}[!ht]
\vspace*{-5mm}
\begin{center}
\includegraphics[angle=0,width=.5\linewidth]{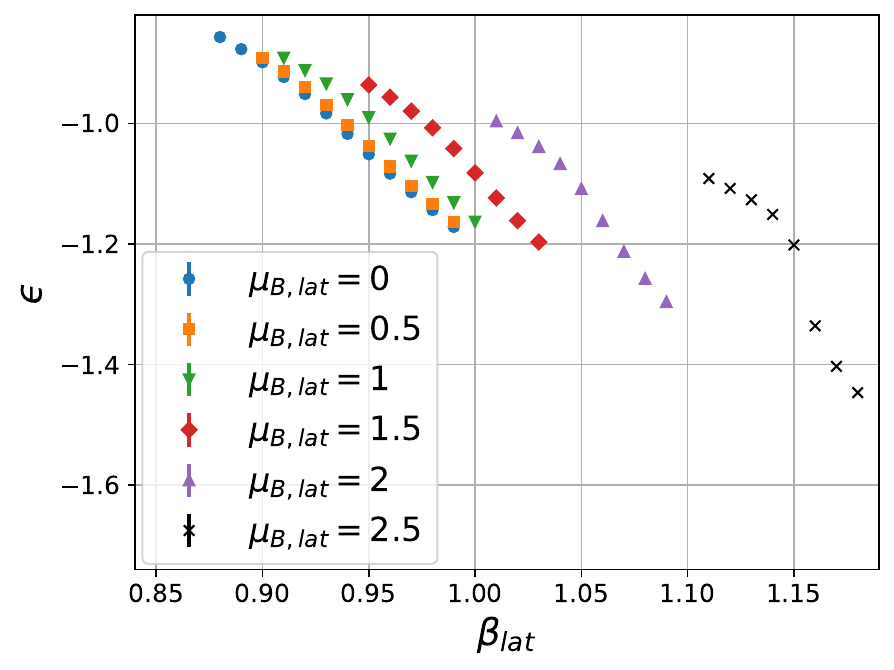}
\includegraphics[angle=0,width=.5\linewidth]{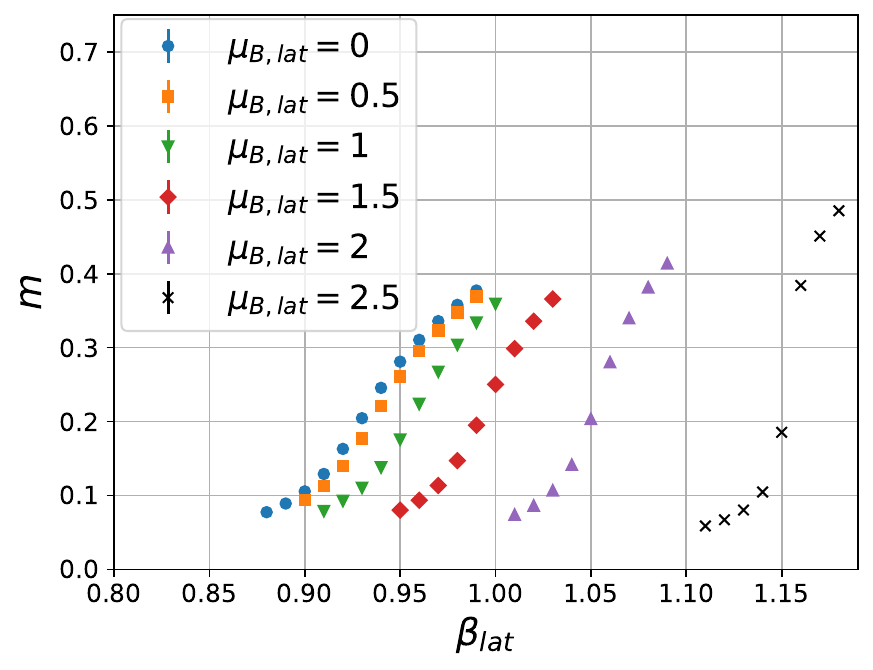}
\includegraphics[angle=0,width=.5\linewidth]{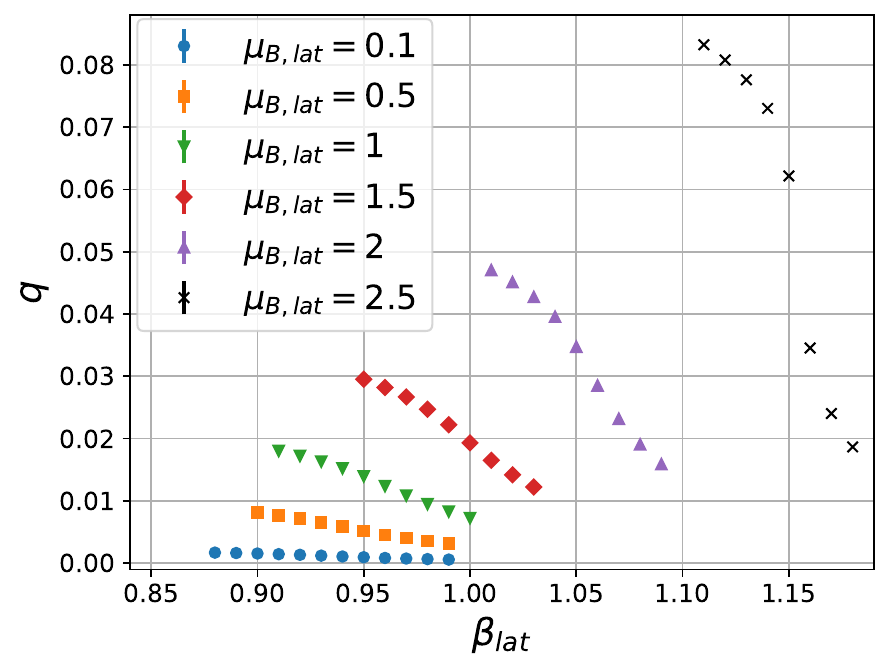}
\end{center}
\vspace*{-6mm}
\caption{Energy density $\epsilon = \la H \ra /V$ (top),
  magnetization density
  $m = \la | \vec M | \ra / V$,  $\vec M = \sum_{x} \vec e_{x}$
  (center) and topological density $q = \la Q \ra /V$ (bottom),
  for $\mu_{B,{\rm lat}} \in [0, 2.5]$, in the lattice volume $V = 20^{3}$.
We see that increasing $\mu_{B,{\rm lat}}$ moves the interval of maximal
slope --- which is the vicinity of the phase transition ---
to larger values of $\beta_{\rm lat}$, and this slope becomes steeper.
In particular at $\mu_{B,{\rm lat}} = 2.5$ one might wonder whether the
phase transition is still of second order, or if this would be a jump
in infinite volume, {\it i.e.}\ a first order phase transition.}
\label{EMagTopDensity}
\vspace*{-15mm}
\end{figure}

Figure \ref{EMagTopDensity} shows three densities, each one
given by a first derivative of the free energy $F$,
\bea
{\rm energy~density} &:& \epsilon = \la H \ra /V
= -\frac{1}{V} \partial_{\beta} (\beta F) \ , \nn \\
{\rm magnetization~density} &:& m = \la | \vec M | \ra / V \ , \quad
\vec M = \sum_{x} \vec e_{x} = \partial_{\vec h} F|_{\vec h = \vec 0} \ , \nn \\
{\rm topological~density}  &:& q = \la Q \ra / V \ , \quad
\la Q \ra = \partial_{\mu_{B}} F \ .
\eea
Note that $m$ is the order parameter.
As we increase $\mu_{B,{\rm lat}} > 0$, positive topological charge is
favored, $\la Q \ra > 0$. In fact, the maximum of $q$ indicated
in Figure \ref{EMagTopDensity} corresponds to an impressive
value of $\la Q \ra \simeq 665.8$.

The enhancement of the topological windings in
the dominant configurations leads to an increase of $\epsilon$, but a
decrease of $m$. This also moves the maximal slope to larger values
of $\beta_{\rm lat}$, which confirms that the critical temperature
decreases as $\mu_{B,{\rm lat}}$ grows.

In particular, at $\mu_{B,{\rm lat}} = 2.5$ all three densities display
a strong slope around $\beta_{\rm lat} \approx 1.15$. At this point,
one might wonder whether it turns into a discontinuous jump in infinite
volume, {\it i.e.}\ whether the first order phase
transition is attained already, or if it is at least nearby.

In order to clarify this question, we move on to quantities
which are given by second-order derivatives of $F$:
the correlation length $\xi$ (given by the exponential decay
of the connected correlation function, which turns into a
cosh-behavior in the finite, periodic volume);
the specific heat (or heat capacity) $c_{V}$;
the magnetic susceptibility\footnote{In numerical studies
of $\chi_{\rm m}$, it is usual to replace
$\la \vec M \ra$ by $\la |\vec M| \ra$, see {\it e.g.}\ Ref.\ \cite{Binder},
since --- strictly speaking --- the former always vanishes in the
absence of an external magnetic field, even in the broken phase,
which prevents us from observing a peak of $\chi_{\rm m}$.
The symbol $\corresp$ (instead of $=$) keeps track of this modification.}
$\chi_{\rm m}$; and the topological susceptibility $\chi_{\rm t}$.
The latter three quantities are defined as
\bea
c_{V} &=& \frac{\beta^{2}}{V} \Big( \la H_{\rm lat}^{2} \ra
- \la H_{\rm lat} \ra^{2} \Big)
 = \frac{\beta^{2}}{V} \partial_{\beta}^{2} (\beta F) \ , \nn \\
\chi_{\rm m} &=&  \frac{\beta}{V} \Big( \la \vec M^{\, 2} \ra -
\la |\vec M| \ra^{2}\Big)
\corresp \frac{1}{V} \partial_{\vec h}^{2} F \ ,
\nn \\
\chi_{\rm t} &=&  \frac{1}{V} \Big( \la Q^{2} \ra -
\la Q \ra^{2} \Big) = \frac{1}{V} \partial_{\mu_{B}}^{2} F \ .
\label{F2}
\eea
As long as the phase transition is of second order, we expect for each
of these three quantities a peak, which would turn into a divergence
at $\beta_{\rm c}$ in the thermodynamic limit $V \to \infty$.

\begin{figure}[h!]
\vspace*{-2mm}
\begin{center}
\includegraphics[angle=0,width=.8\linewidth]{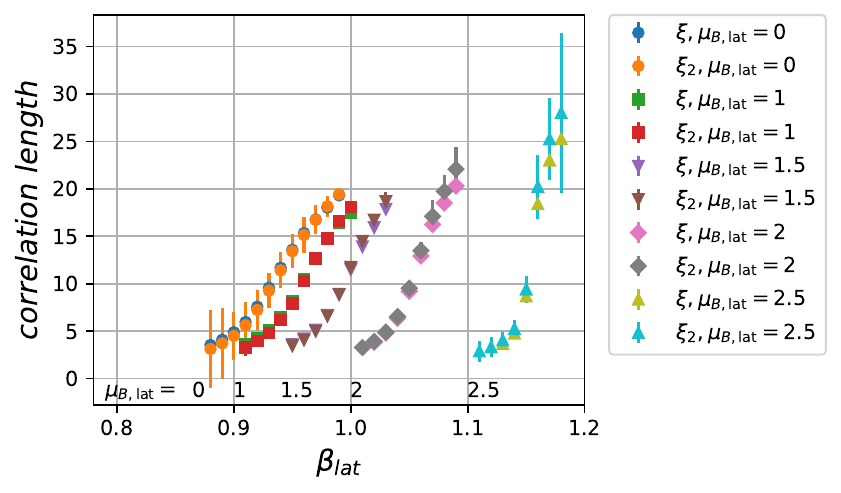}
\end{center}
\vspace*{-6mm}
\caption{Results for the (standard) correlation length $\xi$, and for the
second moment correlation length $\xi_{2}$, in the volume $V=20^{3}$.
The hints for a phase transition of Figure \ref{EMagTopDensity} are
substantiated, now with evidence for the transition to be of second order,
and with consistent estimates for $\beta_{\rm c,lat} (\mu_{B,{\rm lat}})$.}
\label{XiPlot}
\end{figure}

Figure \ref{XiPlot} shows the correlation length $\xi$, at $L=20$.
It is measured with a cosh-fit to the correlation function after mapping
the configurations on one axis. We also show the second moment
correlation length $\xi_{2}$, which is here and in general very similar,
but its numerical measurement is easier since no fit is required,
\bea
\xi_{2} &=& \sqrt{\frac{\tilde G(p = 0) - \tilde G (2\pi/L,0,0)}
{4 \tilde G (2\pi/L,0,0) \sin^{2}(\pi/L)}} \ , \nn \\
\tilde G(p) &=& \frac{1}{V} \sum_{x} \la \vec e_{0} \cdot \vec e_{x}\ra
\exp (- {\rm i} p \cdot x) \ .  \label{xi2eq}
\eea
The maxima in this plot provide another hint for the values of
$\beta_{\rm c,lat}$, consistent with those conjectured from the
maximal slopes in Figure \ref{EMagTopDensity}. However, the
apparent divergence of $\xi$ in the thermodynamic limit is evidence
for the corresponding phase transition to be of second order.

We add that the system only behaves similarly to its infinite-volume limit
when $\xi \ll L$, which prevents us from obtaining precise results for
the critical exponents, see below.

Figure \ref{CvPlots} shows results for the specific heat $c_{V}$.
At $L=20$ (upper plot), we see peaks becoming more pronounced
as $\mu_{B,{\rm lat}}$ increases.
This strongly suggests that the phase transition is
still of second order up to $\mu_{B,{\rm lat}} = 2.5$.

\begin{figure}[h!]
\begin{center}
\includegraphics[angle=0,width=.56\linewidth]{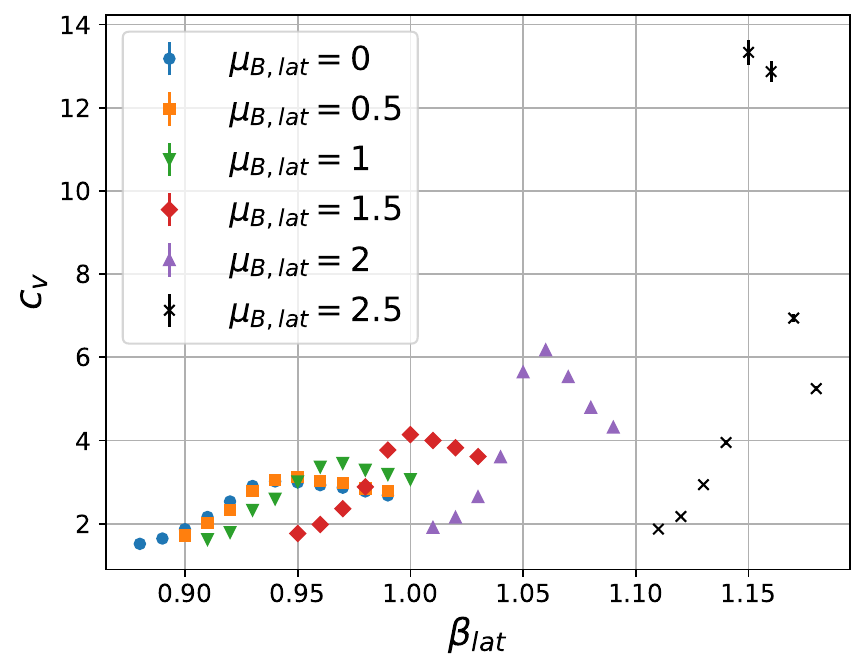}\\
\vspace*{-5.4cm}
\hspace*{1cm}
\includegraphics[angle=0,width=1.25\linewidth]{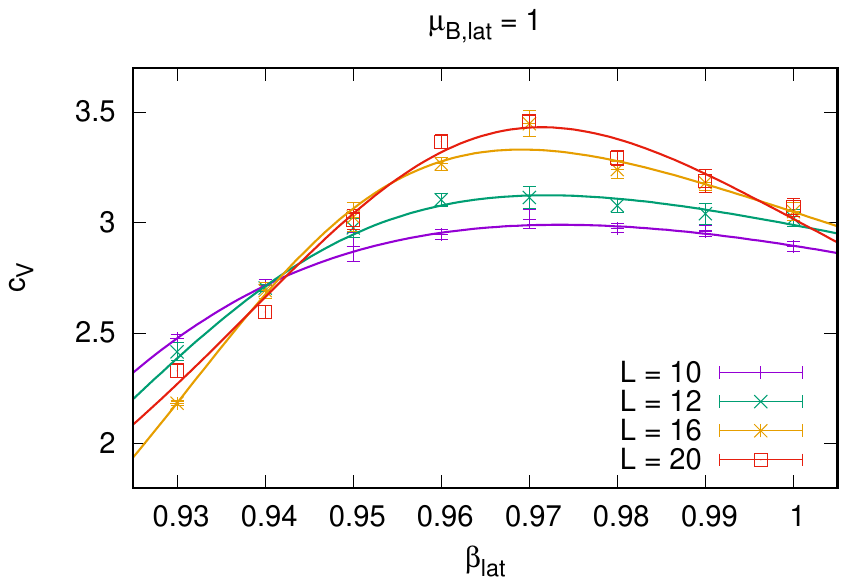}
\vspace*{-2cm}
\end{center}
\vspace*{-6mm}
\caption{Results for the specific heat $c_{V}$.
Above: $\mu_{B,{\rm lat}} = 0, \dots , 2.5$ in the lattice volume
$V = 20^{3}$. At large $\mu_{B,{\rm lat}}$, the peaks are most
pronounced: they confirm that the phase transition is still
of second order, and they are helpful for the determination
of the critical temperature.
Below: Finite-size scaling of the peak at $\mu_{B,{\rm lat}} = 1$
(as an example), fitted with Johnson's $S_U$-function,
see eq.\ (\ref{Johnson}): the value of the maximum, $\beta_{\rm max}$,
hardly moves, and the peak height increases only gradually for larger $L$.}
\label{CvPlots}
\end{figure}

The lower plot compares results for $\mu_{B,{\rm lat}} = 1$ (as an
example) in four volumes. The data are interpolated by Johnson's
$S_U$ (``unbounded system'') function $J$,
\be \label{Johnson}
J(\beta_{\rm lat}) = \frac{c_{1}}{\sqrt{1 +
((\beta_{\rm lat} - c_{2})/c_{3})^{2}}}
\exp \Big( -
\Big[c_{4} + c_{5} \sinh^{-1}((\beta_{\rm lat}-c_{2})/c_{3}) \Big]^{2} \Big)
\ee
with the fitting parameters $c_{1}, \dots , c_{5}$. This provides good
fits over a sizable range in $\beta_{\rm lat}$ for all our values of
$\mu_{B,{\rm lat}}$ (unlike {\it e.g.}\ simple Gaussian fits).
We see that the $\beta_{\rm lat}$-value, where $c_{V}$
takes its maximum, $\beta_{\rm max}$, moves only marginally with $L$.
That property facilitates the thermodynamic extrapolation to
$\beta_{\rm c,lat}$, which we are going to address in Subsection \ref{thermoex}.

We also observe that the peak height increases only
slightly when the volume is enlarged. Finite-size scaling predicts
$c_V(T_{\rm c}) \propto L^{\alpha / \nu}$, where $\alpha$ and $\nu$ are
the critical exponents in the usual notation. This behavior is
consistent with the typically small ratio of these critical exponents:
at $\mu_{B}=0$, Refs.\ \cite{Kanaya,Bielefeld} obtained
$\nu \simeq 0.75$, such that Josephson's scaling law implies
$\alpha / \nu = 3 - 2/\nu \simeq 0.30$. (Unfortunately, the corresponding
values that we extracted at $\mu_{B}>0$ \cite{Edgar,procs} are plagued
by significant finite-size effects.)

\begin{figure}[h!]
\begin{center}
\includegraphics[angle=0,width=.56\linewidth]{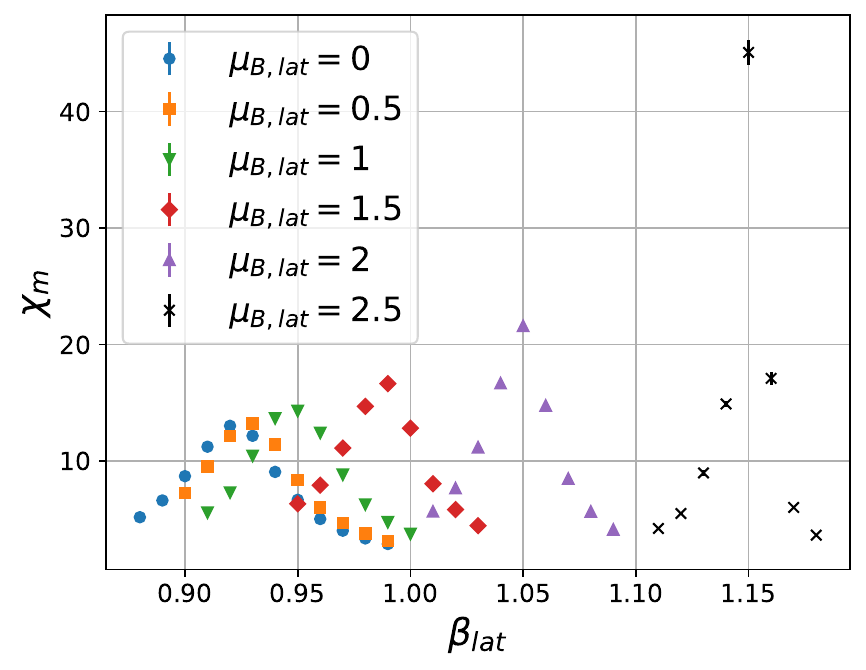} \\
\vspace*{-5.5cm}
\hspace*{1cm}
\includegraphics[angle=0,width=1.25\linewidth]{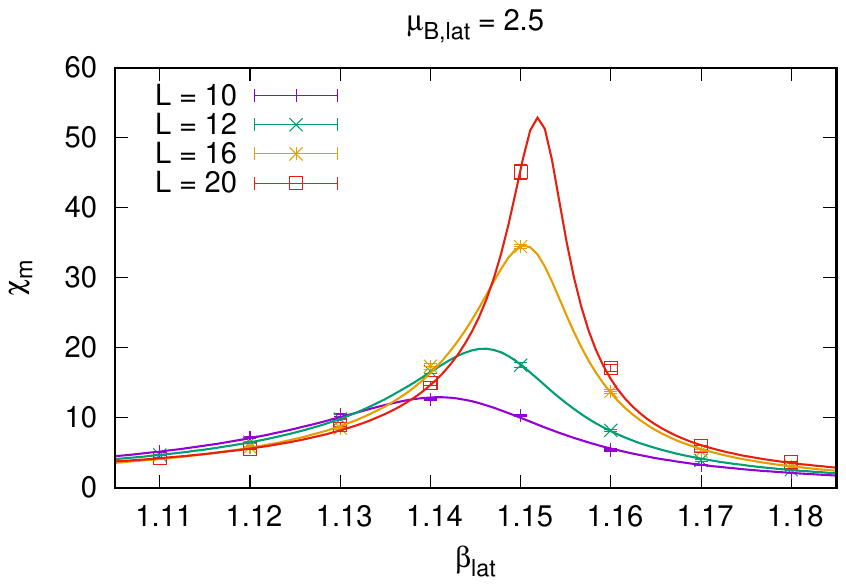}
\vspace*{-2.5cm}
\end{center}
\caption{Results for the magnetic susceptibility $\chi_{\rm m}$,
in analogy to Figure \ref{CvPlots}.
Above: The peaks in $V = 20^{3}$ provide another way to determine
the critical temperature, and to confirm that the phase
transition is of second order, for $\mu_{B,{\rm lat}} \in [0, 2.5]$.
Below: Data obtained in four volumes, again fitted with Johnson's
$S_U$ function of eq.\ (\ref{Johnson}) in order to identify
$\beta_{\rm max}$ by interpolation. Here we show, as an example,
the data for $\mu_{B,{\rm lat}}=2.5$: in this case, we observe a
significant volume-dependence of $\beta_{\rm max}$.
Moreover, the peak height increases substantially when the volume
is enlarged.}
\label{ChimPlots}
\end{figure}

For the magnetic susceptibility $\chi_{\rm m}$, defined in
eq.\ (\ref{F2}), the results are shown in  Figure \ref{ChimPlots}.
Here the peak is more visible even down to $\mu_{B,{\rm lat}} = 0$.
Again it is most pronounced at $\mu_{B,{\rm lat}} = 2.5$, further
affirming that the phase transition is of second order all
over the range investigated here.
In this case, the finite-size effects on $\beta_{\rm max}$ are
stronger than in the case of $c_{V}$.

The strongly increasing peak height in enhanced volume is
consistent with the finite-size scaling relation
$\chi_{\rm m}(T_{\rm c}) \propto L^{\gamma / \nu}$; at $\mu_{B}=0$,
Refs.\ \cite{Kanaya,Bielefeld} measured $\gamma/\nu \simeq 1.97$.
This value, as well as the one for $\nu$ alone, could be determined
from simulations in larger volumes, which were accessible due to
the much shorter auto-correlation time in the absence of a chemical
potential, as shown in Figure \ref{autocorr}.

\begin{figure}[h!]
\begin{center}
\includegraphics[angle=0,width=.5\linewidth]{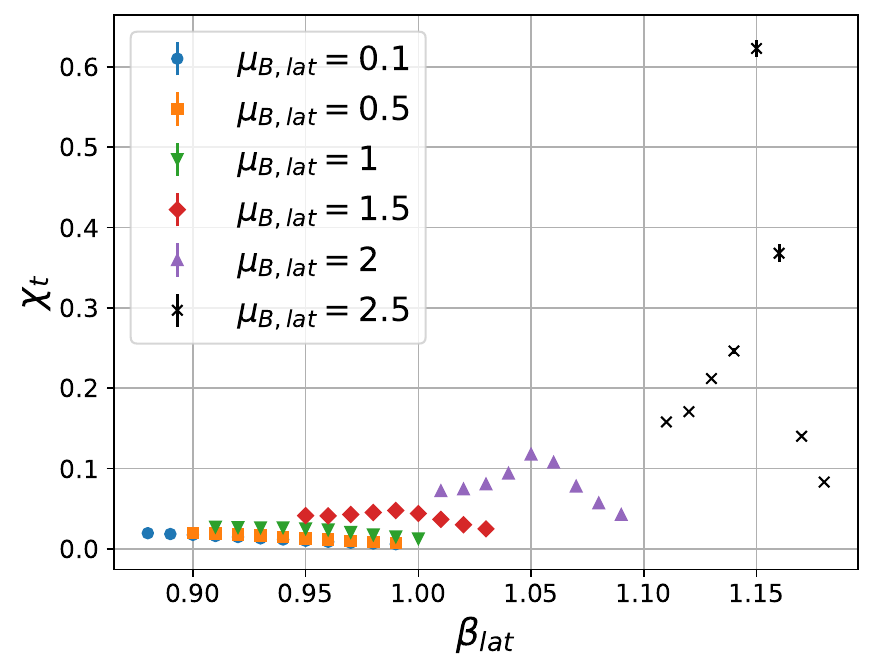}\\
\vspace*{-4.8cm}
\hspace*{1.3cm}
\includegraphics[angle=0,width=1.15\linewidth]{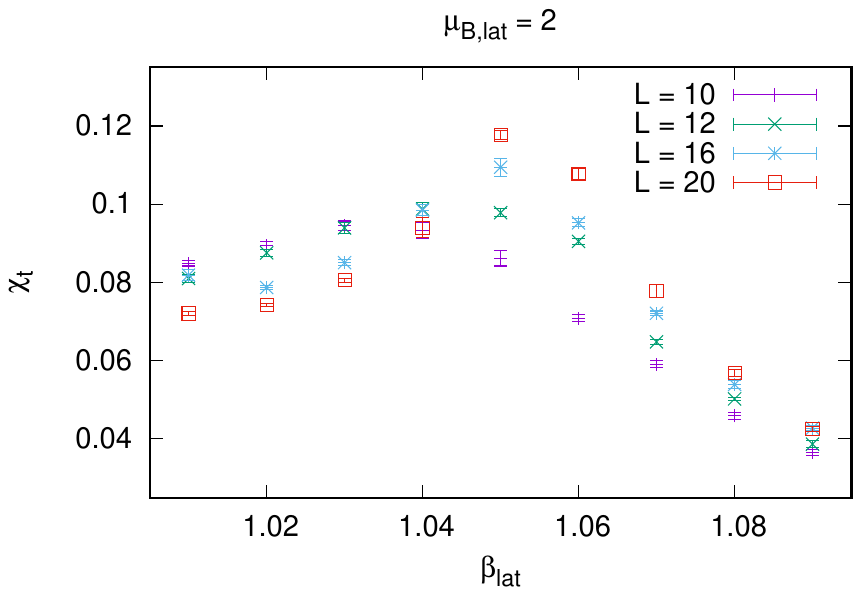}
\end{center}
\vspace*{-1.9cm}
\caption{Results for the topological susceptibility $\chi_{\rm t}$.
Above: $\mu_{B,{\rm lat}} = 0, \dots , 2.5$ in the lattice volume
$V = 20^{3}$. Below: $\mu_{B,{\rm lat}} = 2$ and $V = 10^3, \ 12^3,
\ 16^3$ and $20^3$.}
\label{ChiTop}
\end{figure}

Finally, we proceed to the topological susceptibility $\chi_{\rm t}$,
which is also defined in eq.\ (\ref{F2}). The peaks in
Figure \ref{ChiTop} assure once more that we are dealing with a
second order phase transition for $\mu_{B,{\rm lat}} \leq 2.5$.

Taking the results for $\tau_{H}$,  $\tau_{M}$, $\tau_{Q}$, $\xi$
(and $\xi_{2}$), $c_{V}$, $\chi_{\rm m}$ and $\chi_{\rm t}$
together provides overwhelming evidence that the critical line
extends at least up to $\mu_{B,{\rm lat}} = 2.5$,
despite the strong slopes observed in the densities of
Figure \ref{EMagTopDensity}.

\subsection{Thermodynamic extrapolation}
\label{thermoex}

We have seen indications for the value of the inverse critical
temperature $\beta_{\rm c,lat}$ from various perspectives: the peaks
of the auto-correlation times in Figure \ref{autocorr} and of
the correlation length in Figure \ref{XiPlot}, the
interval of steepest slopes of the densities (as functions of
$\beta_{\rm lat}$) in Figure \ref{EMagTopDensity}, and finally the
peaks of the quantities in eq.\ (\ref{F2}), which are given by
second derivatives of the free energy $F$. The data for the
terms $c_V$ and $\chi_{\rm m}$ are most conclusive for criticality,
and we already denoted the $\beta$-value where we
observe their maxima as $\beta_{\rm max}$. 

\begin{figure}[h!]
\vspace*{-1cm}
\begin{center}
\includegraphics[angle=0,width=.53\linewidth]{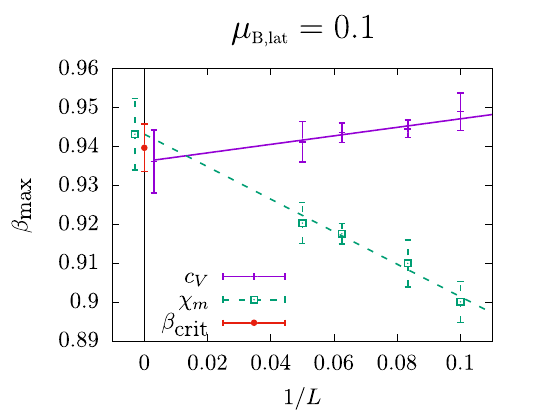}\\
\vspace*{-2mm}
\includegraphics[angle=0,width=.53\linewidth]{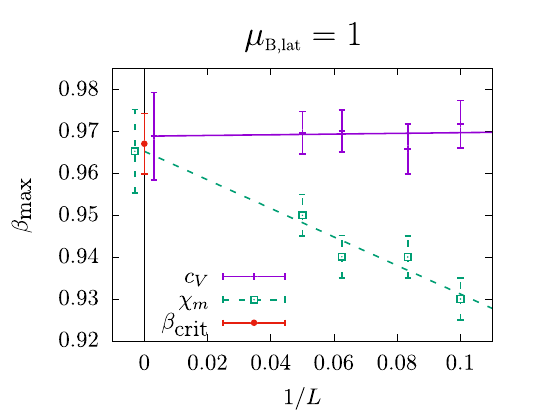}\\
\vspace*{-2mm}
\includegraphics[angle=0,width=.53\linewidth]{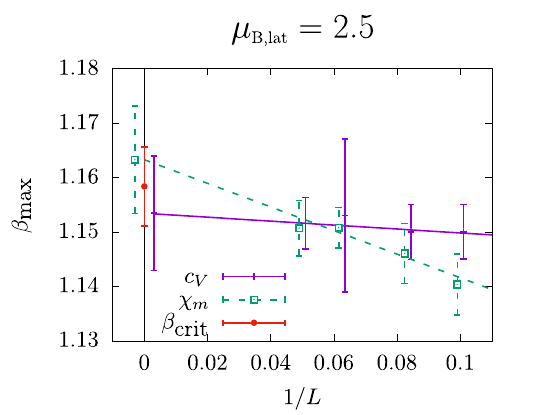}
\end{center}
\vspace*{-6mm}
\caption{Extrapolations of the values of $\beta_{\rm max}$,
obtained by interpolating fits in various volumes, to the critical values
in the thermodynamic limit $\beta_{\rm c,lat}$.
We show results at $\mu_{B,{\rm lat}} =
0.1,\ 1,\ 2.5$, and we see that in each case the extrapolations based on
the specific heat $c_V$ and on the magnetic susceptibility $\chi_{\rm m}$
are consistent.}
\label{extra}
\end{figure}

Figure \ref{extra} shows the extrapolations of $\beta_{\rm max}$ to
the limit $L \to \infty$ for $c_{V}$ and $\chi_{\rm m}$,
at $\mu_{B,{\rm lat}} = 0.1, \ 1,$ and 2.5, as examples.
The extrapolated values are not only consistent with the literature
at $\mu_{B,{\rm lat}}=0$, but they are also in all cases consistent with
each other when deduced from  $c_{V}$ and $\chi_{\rm m}$ (within their
uncertainties). This shows that --- in this regard --- the finite-size
effects are under control thanks to the extrapolations, although the
data are obtained in relatively small volumes. This provides our
results for the inverse critical temperature $\beta_{\rm c,{\rm lat}}$.

Such extrapolations have been performed at each $\mu_{B,{\rm lat}}$-value
in our study, and in all cases we obtained consistent results for
$\beta_{\rm c,lat}(\mu_{B,{\rm lat}})$. These critical values are listed
in Table \ref{betac}. They are the basis for the phase diagram,
as far as we could explore it.

\begin{table}[h!]
 \begin{center}
  \begin{tabular}{|c||c|c|c|c|c|c|}
 \hline
   $\mu_{B,{\rm lat}}$ & 0 & 0.1 & 0.2 & 0.3 & 0.4 & 0.5 \\
 \hline
$\beta_{\rm c,lat}$ & 0.940(5) & 0.940(6) & 9.943(4) & 0.948(7) &
0.941(7) & 0.951(4) \\
\hline
  \end{tabular}
\vspace*{2mm} \\
\begin{tabular}{|c||c|c|c|c|c|c|}
 \hline
  $\mu_{B,{\rm lat}}$ & 0.6 & 0.7 & 0.8 & 0.9 & 1 & 1.1 \\
 \hline
 $\beta_{\rm c,lat}$ & 0.940(7) & 0.948(4) & 0.954(5) &
 0.968(6) & 0.967(7) & 0.974(5) \\
 \hline
  \end{tabular}
\vspace*{2mm} \\
\begin{tabular}{|c||c|c|c|c|c|c|c||c|c|}
 \hline
   $\mu_{B,{\rm lat}}$ & 1.2 & 1.3 & 1.4 & 1.5 & 2 & 2.5 \\
 \hline
   $\beta_{\rm c,lat}$ & 0.986(5) & 0.988(8) & 1.003(8) &
   0.997(8) & 1.064(11) & 1.158(7) \\
 \hline
  \end{tabular}
\end{center}
\vspace*{-2mm}
 \caption{Critical values for the inverse temperature $\beta_{\rm lat}$,
 obtained from extra\-polations to infinite volume, at fixed
 $\mu_{B,{\rm lat}}$, as illustrated (for three examples)
 in Figure \ref{extra}.}
 \label{betac}
 \vspace*{-2mm}
\end{table}

\subsection{Conversion to physical units}

In order to convert the lattice units
to physical units, we need a reference quantity,
which is known in both units. The obvious choice is the critical
temperature $T_{\rm c} = 1 /\beta_{\rm c}$ at $\mu_{B} =0$.
In the 3d O(4) lattice model it has been computed to high
precision by Oevers, who found $\beta_{\rm c,lat} = 0.9359(1)$
\cite{Oevers}, which is compatible with Ref.\ \cite{Kanaya}
and with our result in Table \ref{betac}.
We match this value to the aforementioned critical temperature
obtained in chiral QCD, $T_{\rm c} \simeq 132~{\rm MeV}$ \cite{Ding19},
cf.\ Section \ref{intro}, which leads to
\be
\mu_{B} = \frac{\beta_{\rm c,lat}}{\beta_{\rm c}} \mu_{B,{\rm lat}} \simeq
124~{\rm MeV} ~ \mu_{\rm B,lat} \ .
\ee
Thus we interpret our simulation parameter
$\mu_{\rm B,lat} = 0, \dots , 2.5$ as
\be
0 \leq  \mu_{B} \lesssim 309 ~{\rm MeV} \ .
\ee

\begin{figure}[h!]
\begin{center}
\hspace*{-2mm}   
\includegraphics[angle=0,width=.8\linewidth]{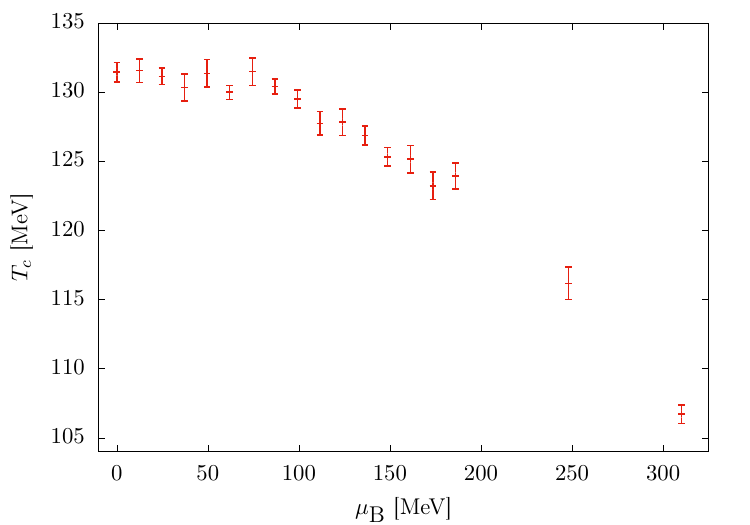}
\end{center}
\vspace*{-6mm}
\caption{The phase diagram for chiral 2-flavor QCD, based on the 3d O(4)
model as an effective theory: our results lead to a critical line
from $T_{\rm c}(0) \simeq 132~{\rm MeV}$ down to
$T_{\rm c}(306~{\rm MeV}) \simeq 106~{\rm MeV}$.}
\label{phasedia}
\end{figure}

By applying the conversion to physical units, we arrive at the
phase diagram in Figure \ref{phasedia}.
This phase diagram summarizes the results of this work, which we
are going to discuss in Section \ref{conclu}.

\section{Conclusions and outlook}

\label{conclu}
Our point of departure was that the 4d O(4) model is most likely
in the universality class of 2-flavor QCD in the chiral
limit, along the lines of Skyrme's theory \cite{Skyrme}.
Unfortunately, O($N$) models cannot cope with more than
two quark flavors, since the pattern of spontaneous symmetry
breaking ${\rm O}(N) \to {\rm O}(N-1)$ is not isomorphic to
${\rm SU}(N_{\rm f}) \otimes {\rm SU}(N_{\rm f}) \to
{\rm SU}(N_{\rm f})$ for any $N_{\rm f} > 2$ \cite{WB}.

Next we assumed the temperature to be high enough for a
dimensional reduction to be a sensible approximation.
This leads to the 3d O(4) model, where the configurations carry
topological charge $Q$, which can be identified with the baryon
number $B$.
Hence, the baryon chemical potential $\mu_{B}$ takes the form of an
imaginary $\theta$-vacuum angle. This term does not cause
any sign problem in Monte Carlo simulations.

We simulated this model with a cluster algorithm
and monitored the critical line in the range of
$\mu_{B,{\rm lat}} = 0, \dots , 2.5$ (in lattice units).
In physical units, the maximal value is estimated as
$\mu_{B} \simeq 309~{\rm MeV}$.

The critical temperature $T_{\rm c}(\mu_{B})$ decreases monotonically,
see Figure \ref{phasedia},
in agreement with the generally expected behavior in QCD.
In our study, it moves from $T_{\rm c}(0) \simeq 132~{\rm MeV}$
down to $T_{\rm c}(309~{\rm MeV}) \simeq 106~{\rm MeV}$.
In this range, we consistently observed a second order phase
transition, but in Figure \ref{EMagTopDensity}
there are hints for the final point to be near
a crossing to a first order phase transition.

\begin{figure}[h!]
\vspace*{-8mm}
\begin{center}
\includegraphics[angle=0,width=1.\linewidth]{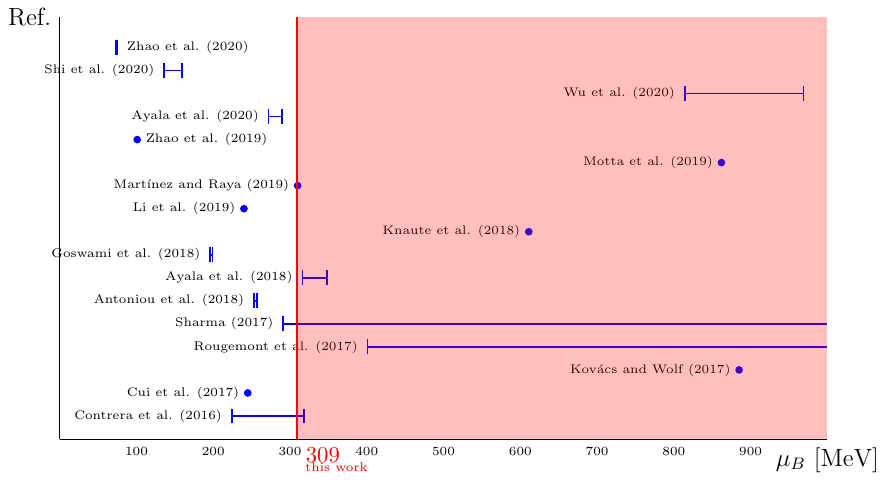}
\vspace*{3mm} \\
\includegraphics[angle=0,width=.7\linewidth]{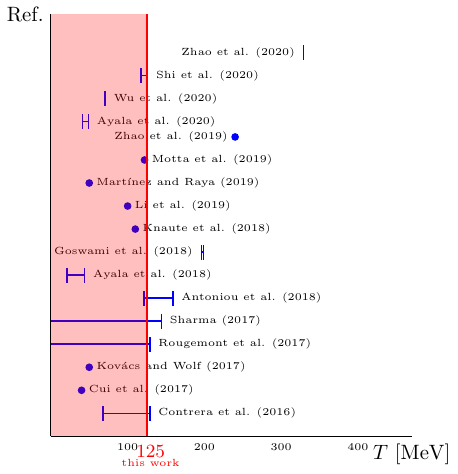}
\end{center}
\vspace*{-0.5cm}
\caption{Overview of estimates for the location of the CEP
in the $(\mu_{B},T)$-plane. The conjectures in 17 papers ---
based on a broad variety of assumptions and techniques ---
are compared to the outcome of this work.}
\label{overview}
\vspace*{-1cm}
\end{figure}

If we trust this effective theory, then our results imply bounds
on the location of a (possible) Critical Endpoint:
$T_{\rm CEP} < 106~{\rm MeV}$, $\mu_{B, {\rm CEP}} > 309~{\rm MeV}$,
which agrees with part of the predictions in the literature.
An overview over such predictions is given in Figure \ref{overview},
and the corresponding bibliography is collected in Ref.\
\cite{CEPlit}. Part of this list is adopted from Ref.\ \cite{Ayala18}.

The inclusion of quark masses corresponds --- in our effective theory
--- to the addition of a constant, external magnetic field.
In this modified model, we observe a cross-over rather than a phase
transition \cite{JAGH,procs}. This alleviates the problem
with the very long auto-correlation times in the current study
(cf.\ Subsection \ref{simu}), and the systematic errors due to
finite-size effects (cf.\ Subsection \ref{numires}),
since neither the autocorrelation time nor
the correlation length diverge at the cross-over.
On the other hand, the pseudo-critical value $\beta_{\rm pc}$
is somewhat ambiguous, depending on the criterion that one
considers. This scenario is under investigation \cite{prep};
preliminary results do again not reveal a Critical Endpoint up to
$\mu_{B} \simeq 250~{\rm MeV}$ \cite{JAGH,procs}.

We are convinced that our phase diagram in Figure \ref{phasedia}
provides a noteworthy orientation: the shape of the critical line is
compatible with the literature, and it allows us to conjecture bounds
for the CEP.
 \newline

\noindent
{\bf Acknowledgments:} We are indebted to Uwe-Jens Wiese for
inspiring this project. We also thank Arturo Fern\'{a}ndez T\'{e}llez
and Miguel \'{A}ngel Nava Blanco for their contributions at an early
stage, Alexis Aguilar-Arevalo for advice with the data analysis,
and Victor Mu\~{n}oz-Vitelly for reading the manuscript.
The simulations were performed on the cluster of the Instituto
de Ciencias Nucleares; we thank Luciano D\'{\i}az Gonz\'{a}lez
and Eduardo Murrieta Le\'{o}n for technical assistance.
This project was supported by UNAM-DGAPA through the PAPIIT projects
IG100219, ``Exploraci\'{o}n te\'{o}rica y experimental del diagrama
de fase de la cromodin\'{a}mica cu\'{a}ntica'' and IG100322,
``Materia fuertemente acoplada en condiciones extremas con el MPD-NICA'',
and by the Consejo Nacional de Humanidades, Ciencia y Tecnolog\'{\i}a
(CONAHCYT), which has now been converted into the
Secretar\'{\i}a de Ciencia, Humanidades, Tecnolog\'{\i}a e
Innovaci\'{o}n (SECIHTI).

\end{document}